\documentclass[prd,amssymb,aps,twocolumn,epsfig]{revtex4}

\usepackage{graphicx}
\usepackage{dcolumn}   
\usepackage{amsmath,amssymb}
\usepackage{bm}
\usepackage[colorlinks,
              linkcolor=blue,
              citecolor=blue,
              urlcolor=blue,
              anchorcolor=blue]{hyperref}

\def\be{\begin{equation}}       \def\ee{\end{equation}}
\def\bea{\begin{eqnarray}}      \def\eea{\end{eqnarray}}
\def\ba{\begin{array} }
\def\ea{\end{array} }
\def\bc{\begin{center}}
\def\ec{\end{center}}
\def\bnum{\begin{enumerate} }
\def\enum{\end{enumerate}}

\def\=>{\Rightarrow}
\def\>{\rightarrow}

\date{\today}

\begin{document}

\title{Probing gauge-phobic heavy Higgs bosons at high energy hadron colliders}


\author{Yu-Ping Kuang$^{1,2}$\footnote{ypkuang@mail.tsinghua.edu.cn} and Ling-Hao Xia$^{1}$\footnote{xlh10@mails.tsinghua.edu.cn}}

\affiliation{$^1$ Department of Physics, Tsinghua
University,Beijing, 100084, China} \affiliation{$^2$ Center for
High Energy Physics, Tsinghua University, Beijing, 100084, China}
\begin{abstract}
\null\noindent{\bf Abstract}

We study the probe of the gauge-phobic (or nearly gauge-phobic)
heavy Higgs bosons (GPHB) at high energy hadron colliders
including the 14 TeV LHC and the 50 TeV Super Proton-Proton
Collider (SppC). We take the process $pp\to t\bar t t\bar t$, and
study it at the hadron level including simulating the jet
formation and top quark tagging (with jet substructure). We show
that, for a GPHB with $M^{}_H<800$ GeV, $M^{}_H$ can be determined
by adjusting the value of $M^{}_H$ in the theoretical
$p^{}_T(b_1)$ distribution to fit the observed $p^{}_T(b_1)$
distribution, and the resonance peak can be seen at the SppC for
$M^{}_H$=800 GeV and 1
TeV. \\
 \null\noindent{PACS numbers: 14.80.Ec, 13.90.+i, 14.65.Ha}
\end{abstract}
\null\noindent{\null\hspace{5cm}TUHEP-TH-15182}

\maketitle
\null\noindent{\bf I. Introduction}

 The most important event in the 7 and 8 TeV
runs of the LHC is the discovery of the 125 GeV Higgs boson
\cite{ATLAS-CMS12}. The ATLAS and CMS collaborations have been
making efforts to measure its couplings to the gauge bosons,
$\tau^+\tau^-$, and $b\bar b$
\cite{ATLAS-higgs-measurement}\cite{CMS-higgs-measurement}\cite{Flechl},
and the obtained results are all consistent with the corresponding
standard model (SM) couplings to the present experimental
precision. However, this does not imply that the SM is the final
theory of fundamental interactions since the SM with a 125 GeV
Higgs boson suffers from various shortcomings, such as the
well-known theoretical problems
\cite{triviality}\cite{unnaturalness}\cite{vacuum-instability};
the facts that it does not include the dark matter; it can neither
predict the mass of the Higgs boson nor predict the masses of all
the fermions, etc. Searching for new physics beyond the SM is the
most important goal of future particle physics studies. So far
there is no evidence of some well-known new physics models such as
supersymmetry, large extra dimensions, etc. We know that most
known new physics models contain more than one Higgs bosons in
which the lightest one may be very close to the SM Higgs boson,
and the masses of other heavy Higgs bosons are usually of the
order of $10^2\--10^3$ GeV. So that the discovered 125 GeV Higgs
boson may be the lightest Higgs boson in certain new physics
model, and probing other heavy Higgs bosons may be a feasible way
of searching for new physics. Heavy Higgs bosons in the minimal
supersymmetric extension of the SM (MSSM) and the two-Higgs-doulet
model (2HDM) have been searched for at the LHC with negative
results \cite{ATLAS-PRD89}. So that performing a model-independent
search is more effective.

In our previous paper \cite{KRX-PRD14}, we proposed a sensitive
way of probing anomalous heavy neutral Higgs bosons $H$
model-independently at the 14 TeV LHC via the process $pp\to
VH^\ast\to VVV\to \ell^+\nu^{}_\ell j^{}_1j^{}_2j^{}_3j^{}_4$,
where $V=W,Z$. We showed that the resonance peak of $H$ and the
values of the anomalous coupling constants $f^{}_W,f^{}_{WW}$ can
be measured experimentally provided the $HVV$ couplings are not so
small. Of course, this can not be applied to gauge-phobic (or
nearly gauge-phobic) heavy Higgs bosons (GPHB) with vanishing (or
very small) $HVV$ couplings. For a GPHB $H$ with mass $M^{}_H$ in
the 400 GeV to a few TeV range, its main decay mode is $H\to t\bar
t$. If we simply consider the $t\bar t$ final state, the
background will be extremely large. Ref.\,\cite{BhupalDev14}
showed that the process $pp\to t\bar t H\to t\bar t t\bar t$ is a
feasible process, and studied it at the parton level in the 2HDM.
In this paper, we take the process $pp\to t\bar t t\bar t$, and
study the full tree level contribution at the hadron level
including simulating the jet formation and top quark tagging (with
jet substructure). We show that, with suitable kinematic cuts, the
GPHB mass $M^{}_H$ can be determined by adjusting the value of
$M^{}_H$ in the theoretical $p^{}_T(b_1)$ distribution to fit the
experimentally measured $p^{}_T(b_1)$ distribution for
$M^{}_H<800$ GeV at both the LHC and the 50 TeV Super
Proton-Proton Collider (SppC) considered in Beijing, and the
resonance peak can be seen at the SppC for $M^{}_H$=800 GeV and 1
TeV.

\null\noindent{\bf II. Calculation and Results}

The general Yukawa coupling of the GPHB with
the top quark can be written as
\begin{equation}                         
y^H_t \bar t \Phi_H t \equiv C^{}_t y^{SM}_t \bar t \Phi_H t,
\label{Yukawa}
\end{equation}
where $C_t$ is a parameter reflecting the deviation from the SM
coupling. For simplicity, we take $C^{}_t\approx 1$. In this
paper, we take $M^{}_H=$ 400 GeV, 600 GeV, 800 GeV, and 1 TeV as
examples to do the calculation. \null\vspace{-0cm}
\begin{table}[h]                                   
\label{acceptance} \tabcolsep 12pt \caption{Detector acceptance
according to DELPHES3}
\begin{tabular}{ccccc}
\hline\hline
 &$\mu$&$e$&{\rm jet}&{\rm photon}\\
 \hline\\
 $|\eta|_{max}$~~~~~~~~~~&2.4&2.5&5&2.5\\
 $p^{}_{Tmax}$(GeV)&10&10&20&0.5\\
 \hline\hline
 \end{tabular}
 \label{acceptance}
 \end{table}

Ref.\,\cite{four-top-NLO} shows that the next-to-leading-order
correction to the four-top production cross section in the SM is
not so large. So we do the leading full tree level simulation of
$pp \rightarrow t\bar{t}t\bar{t}$ at hadron colliders in this
paper. We use MadGraph5\cite{madgraph5}, FeynRules\cite{feynrules}
and Pythia6.4\cite{pythia-6.4} to simulate the signals and the
backgrounds. We take CTEQ6.1\cite{cteq6.1} as the parton
distribution function (PDF). Delphes3 \cite{delphes3} and fastjet
\cite{fastjet} are used to simulate detector acceptance and jet
reconstruction. Our detector acceptance is shown in Table
\ref{acceptance} referring to the design of CMS detector
\cite{CMS-detector}.

\begin{figure}[htbp]                               
\includegraphics[width=0.4\textwidth]{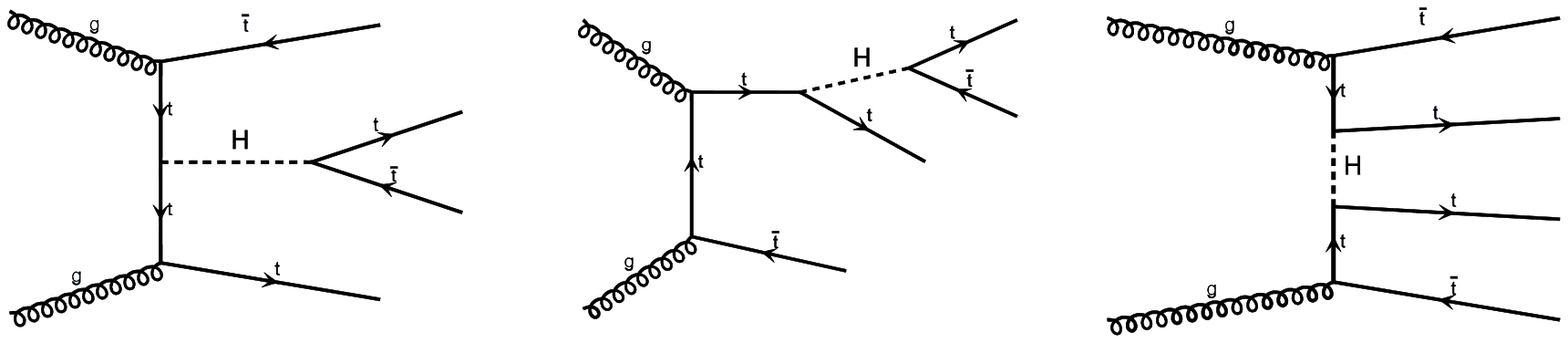}\\
\centerline{(a)}
\includegraphics[width=0.25\textwidth]{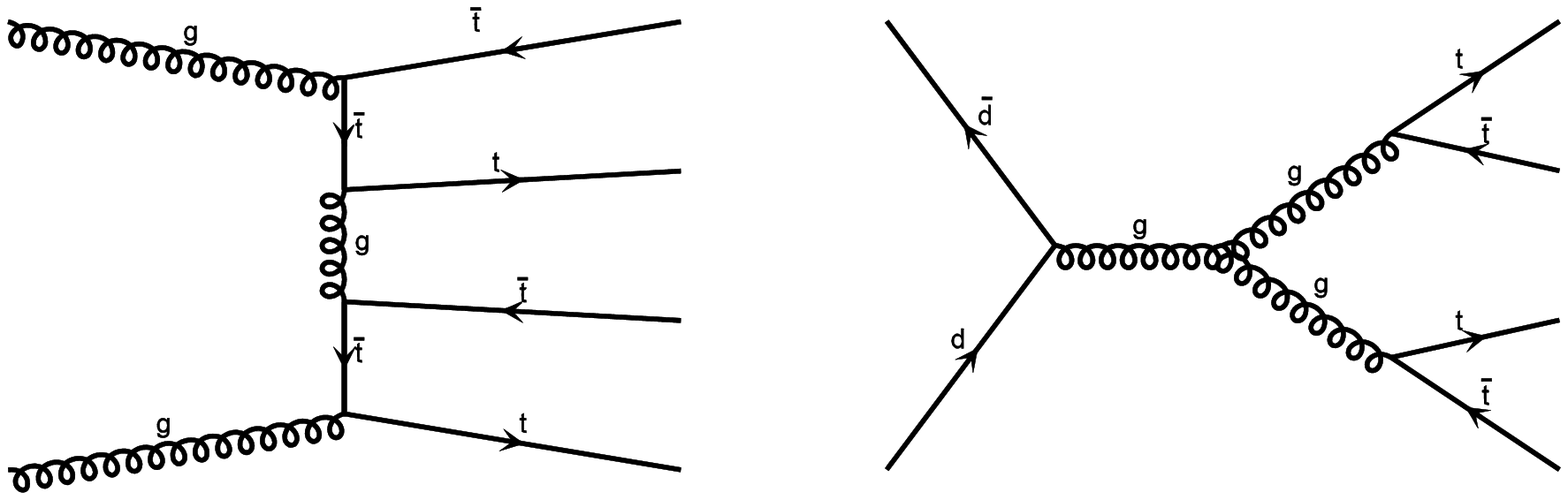}\\
\centerline{(b)} \caption{Some typical Feynmann diagrams in $pp
\rightarrow
 t\bar{t}t\bar{t}$: (a) for S, and (b) for
IB.}
\label{higgs-top-pair}
\end{figure}

Some typical Feynman diagrams for the signal (S) and the
irreducible background (IB) in $pp \rightarrow t\bar{t}t\bar{t}$
are depicted in Fig.\,\ref{higgs-top-pair}. These two amplitudes
will interfere with each other, so that they should be calculated
together.

To suppress the SM background, we need some of the decay modes of
the top quarks to include leptons. The CMS collaboration has
studied the production of $t\bar tt\bar t$ in the SM with the
final state including a single lepton and multiple jets at the 8
TeV LHC \cite{CMS-TOP-13-012}, and has shown that it suffers from
a very strong SM background. The ATLAS Collaboration analyzed the
production of $t\bar tt\bar t$, and  pointed out that, with the
energy and integrated luminosity of the LHC Run I, the most
favorable final state is the one with two same-sign leptons
\cite{ATLAS-4top}. Processes with fewer final state leptons can
have larger cross sections but with lower signal to background
ratios. In this paper, we shall present our simulation results for
two kinds of final states. First we study the final states
including three charged leptons (3-lepton mode) in which the SM
reducible backgrounds (RBs) are highly suppressed. Next we study
the final states including two opposite-sign leptons
 (2-lepton mode) in which the resonance peak of $H$ can be
observed.

\null\noindent{\bf A. The case of 3-lepton mode}

\begin{figure}[h]                            
\includegraphics[width=0.4\textwidth]{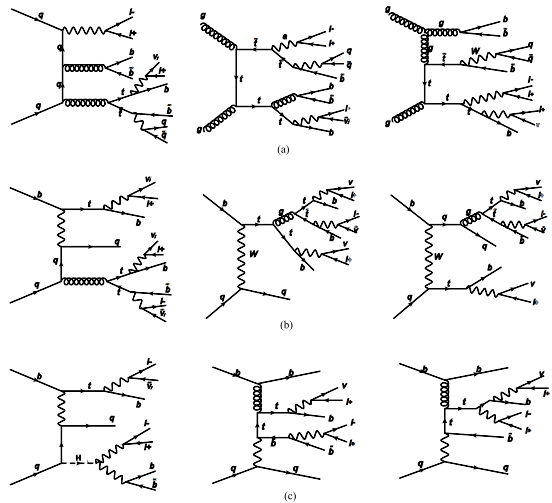}
\null\vspace{-0.2cm} \caption{Fenmann diagrams for typical RBs in
the case of 3-lepton mode.}
\label{triple-leptons}
\end{figure}

We first study the case of 3-lepton mode, i.e. three of the four
top (anti-top) quarks in the final state decay semileptonically
while the other top (anti-top) quark decays hadronically. A signal
event must contain $2\ell^+ 1\ell^-$ or $1\ell^+ 2\ell^-$ ($\ell$
denotes $\mu$ or $e$), and at least three tagged b jets. For
reducible backgrounds (RB), there are
three most important processes which can mimic the signal:\\
\null\noindent $\bm\bullet$ {\bf RB1: $\bm{\ell^+ \ell^-
b}\bar{\bm{b}}\bm{ t}\bar{\bm{t}}$}\\
 Processes with $\ell^+
\ell^-$, $b\bar{b}$, and $t\bar{t}$ in the final state. One $t$
(or $\bar t$) decays semileptonically and the
other $\bar t$ (or $t$) decays hadronically (cf. Fig.\, \ref{triple-leptons}(a)).\\
\null\noindent {\bf $\bm\bullet$ RB2: $\bm{t}\bar{\bm
t}\bm{t}\bm{+{\rm jets}}$ and $\bm{t}\bar{\bm t}\bar{t}\bm{+{\rm jets}}$}\\
Processes with $t\bar{t}$ and another single top (anti-top) quark
with extra jets in the final state. All the three top (anti-top)
quarks decay semileptonically. We take one of the extra jets
generated by matrix element and the other
jets generated by parton shower (cf. Fig.\,\ref{triple-leptons}(b)).\\
\null\noindent {\bf$\bm\bullet$ RB3: $\bm{\ell^+ \ell^- b}\bar{\bm
b}\bm{t}\bm{+{\rm jets}}$ and $\bm{\ell^+ \ell^- b}\bar{\bm
b}\bar{\bm
t}\bm{+{\rm jets}}$}\\
 Processes with $\ell^+\ell^-$, $b\bar{b}$, and a top
(anti-top) quark with extra jets in the final state. The top
(anti-top) quark decays semileptonically. We take one of the extra
jets generated by matrix element and the other jets generated by
parton shower (cf. FIG.\, \ref{triple-leptons}(c)).

We generate events for S+IB, IB and RBs at the 14 TeV LHC and the
50 TeV SppC, and do the detector simulation with Delphes3
\cite{delphes3}. Anti-$k_T$ algorithm with $R=0.5$ are used to
cluster jets and tag b jets. For the b-tagging efficiencies, we
use the values in Ref.\, \cite{snowmass}. Then we apply the
following kinematic cut: Requiring the event to contain three
leptons ($2\ell^+ 1\ell^-$ or $2\ell^- 1\ell^+$) and at least
three tagged b jets, i.e.
\begin{eqnarray}                       
\label{triple-leptons-cut}
N(\ell^\pm)=2,~N(\ell^\mp)=1,~~N(b)\geq 3.
\end{eqnarray}

Since the S and the IB cross sections cannot be exactly separated,
we define the IB, S, and total-background (B) cross section by
$\sigma^{}_{\rm{IB}}\equiv
\sigma^{}_{\rm{S+IB}}(C_t=0)$, $\sigma^{}_{\rm{ S}}\equiv
\sigma^{}_{\rm{S+IB}}(C_t\ne 0)-\sigma^{}_{\rm{IB}}$
\cite{footnote-1}, and $\sigma^{}_{\rm{B}}\equiv
\sigma^{}_{\rm{IB}}+\sigma^{}_{\rm{RB}}$, respectively. Here
$\sigma^{}_{\rm B}$ is the total SM background.

After this cut, the cross sections of S+IB, IB and RBs for various
values of $M^{}_H$ are shown in Table~\ref{3Lefficiency}. We see
that RBs are negligibly small after the cut, and a smaller
$M^{}_H$ leads to larger signal cross section. Since the RBs are
already negligible, we do not need to impose the top quark tagging
requirement taking account of the jet substructure.
\begin{widetext}

\begin{table}[h]                           
\begin{small}
\tabcolsep 9pt \caption{\label{3Lefficiency}Cross sections (in fb)
of S+IB, IB and RBs for various values of $M^{}_H$ after the cut
in case of 3-lepton mode.}
\begin{tabular}{ccccccccc}
\hline \hline &&S+IB&&&IB&RB1&RB2&RB3\\
            & 400 GeV & 600 GeV & 800 GeV & 1000 GeV &  &  & &  \\
\hline
14 TeV LHC   &   0.14  &   0.10  &  0.079   & 0.066    &  0.060   &   $1.3\times10^{-6}$  &  $1.3\times10^{-9}$  & $6.4\times10^{-7}$ \\
50 TeV SppC  &   3.87   &   3.31   &  2.63     & 2.21      &  1.92    &   0.039               &  0.00032             & 0.0039             \\
\hline \hline
\end{tabular}
\end{small}
\end{table}
\end{widetext}

For an integrated luminosity ${\sf L}_{\rm {int}}$, the signal and
background event numbers are $N^{}_{\rm S}={\sf
L}^{}_{\rm{int}}\sigma^{}_{\rm {S}}$ and $N^{}_{\rm{ B}}={\sf
L}^{}_{\rm{int}}\sigma^{}_{{\rm B}}$, respectively. For large
enough $N^{}_{\rm{ S}}$ and $N^{}_{\rm {B}}$, the statistical
significance is defined as $\sigma^{}_{\rm{stat}}=N^{}_{\rm
{S}}/\sqrt{N^{}_{\rm{ B}}}$. The ${\sf L}^{}_{\rm{int}}$ needed
for $1\sigma$, $3\sigma$, and $5\sigma$ deviations at the 14 TeV
LHC and the 50 TeV SppC are shown in Table\,
\ref{tripel-leptons-luminosity}. \null\vspace{-0.5cm}
\begin{table}[h]                                
\begin{center}
\caption{\label{tripel-leptons-luminosity} Integrated luminosity
(in fb$^{-1}$) needed for $1\sigma,3\sigma$ and $5\sigma$
deviations at the 14 TeV LHC and the 50 TeV SppC for different
values of $M^{}_H$.}
\begin{tabular}{cccccc}
\hline \hline
           &            & 400 GeV & 600 GeV & 800 GeV & 1000 GeV \\
\hline
           &$1\sigma$   &10       &37      &172      &1664      \\
14 TeV LHC   &$3\sigma$   &87       &341     &1552     &14977     \\
           &$5\sigma$   &241      &947     &4312     &41604    \\
\hline
           &$1\sigma$   &0.52     &1.0     &3.8      &23       \\
50 TeV SppC  &$3\sigma$   &4.6      &9.1     &34       &205      \\
           &$5\sigma$   &13       &25      &96       &570      \\
\hline \hline
\end{tabular}
\end{center}
\end{table}
We see that the $5\sigma$ case for $M^{}_H=600$ GeV, and the
$3\sigma$ and $5\sigma$ cases for $M^{}_H\ge 800$ GeV at the LHC
require very high luminosities which need the upgraded high
luminosity LHC. All other cases can be reached at the present 14
TeV LHC and the SppC.

\begin{figure}[h]                        
\includegraphics[width=0.5\textwidth]{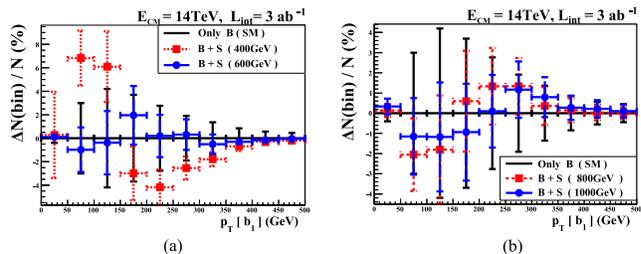}
\null\vspace{-0.8cm} \caption{$\Delta N^{}_{\rm
S+B}(\rm{bin})/N^{}_{\rm S+B}-\Delta N^{}_{\rm
B}(\rm{bin})/N^{}_{\rm B}$ with statistical errors at 14 TeV LHC:
(a) $M^{}_H$=400 and 600 GeV, (b) $M^{}_H$=800 and 1000 GeV.}
\label{LHC-p_T}
\end{figure}

\begin{figure}[h]                            
\includegraphics[width=0.5\textwidth]{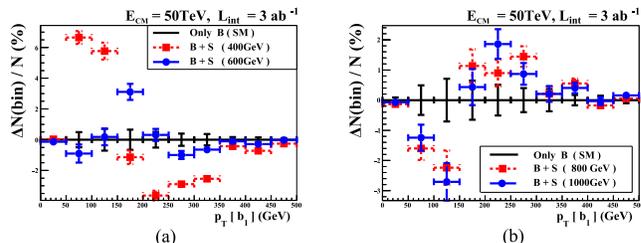}
\null\vspace{-0.8cm} \caption{$\Delta N^{}_{\rm
S+B}(\rm{bin})/N^{}_{\rm S+B}-\Delta N^{}_{\rm
B}(\rm{bin})/N^{}_{\rm B}$ with statistical errors at 50 TeV SppC:
(a) $M^{}_H$=400 and 600 GeV, (b) $M^{}_H$=800 and 1000 GeV.}
\label{SppC-p_T}
\end{figure}

Since there are three missing neutrinos in the case of 3-lepton
mode, it is hard to construct the invariant mass distribution of
the top quark pair from the decay of $H$ to show the resonance
peak of $H$. However, the values of $M^{}_H$ may affect the
distribution of certain kinematic observable from which we may
determine the value of $M^{}_H$. In the final state, the $b$
($\bar b$) quark is the secondary decay product which is more
closely related to $H$ than the charged leptons and ordinary jets
(from $W$ decays) do. Let $b_1$ be the $b$ quark with largest
$p^{}_T$. We choose the $p^{}_T(b_1)$ distribution to reflect the
effects of different values of $M^{}_H$. Note that the unknown
value of $C_t$ also affects the $p^{}_T(b_1)$ distribution as an
overall factor. To eliminate this effect, we separate the
$p^{}_T(b_1)$ axis into a certain number of bins. Let $\Delta
N(\rm{bin})$ be the number of events within a bin at a certain
value of $p^{}_T(b_1)$, and $N\equiv \sum_{\rm{bin}} \Delta
N(\rm{bin})$ be the total number of events in the $p^{}_T(b_1)$
distribution. We then take the {\it normalized} distribution (ND),
$\Delta N(\rm{bin})/N$, for both S+B (in which the unknown
$C^{}_t$ dependence is cancelled) and B (SM background). In
Figs.\,\ref{LHC-p_T} and \ref{SppC-p_T}, we plot the ND, $\Delta
N^{}_{\rm S+B}(\rm{bin})/N^{}_{\rm S+B}-\Delta N^{}_{\rm
B}(\rm{bin})/N^{}_{\rm B}$, with statistical errors for various
values of $M^{}_H$ with an integrated luminosity of 3 ab$^{-1}$ at
the 14 TeV LHC and the 50 TeV SppC, respectively.

We see that the distributions are clearly distinguishable for
$M^{}_H<800$ GeV at the LHC (the distributions for $M^{}_H>800$
GeV can hardly be distinguished at the 14 TeV LHC), and they are
distinguishable for all values of $M^{}_H$ at the 50 TeV SppC.
Thus the value of $M^{}_H$ can be determined by adjusting the
value of $M^{}_H$ in the theoretical $p^{}_T(b_1)$ distribution to
fit the experimentally measured $p^{}_T(b_1)$ distribution.

Furthermore, if we adjust the value of $C^{}_t$ in $\sigma^{}_{\rm
S+B}$ to fit the observed cross section, we may also determine the
value of $C^{}_t$ for the GPHB in nature. Of course the
uncertainty depends on the experimental error.

 \null\noindent{\bf
B. The case of 2-lepton mode}\\
\begin{figure}[h]                           
\includegraphics[width=0.38\textwidth]{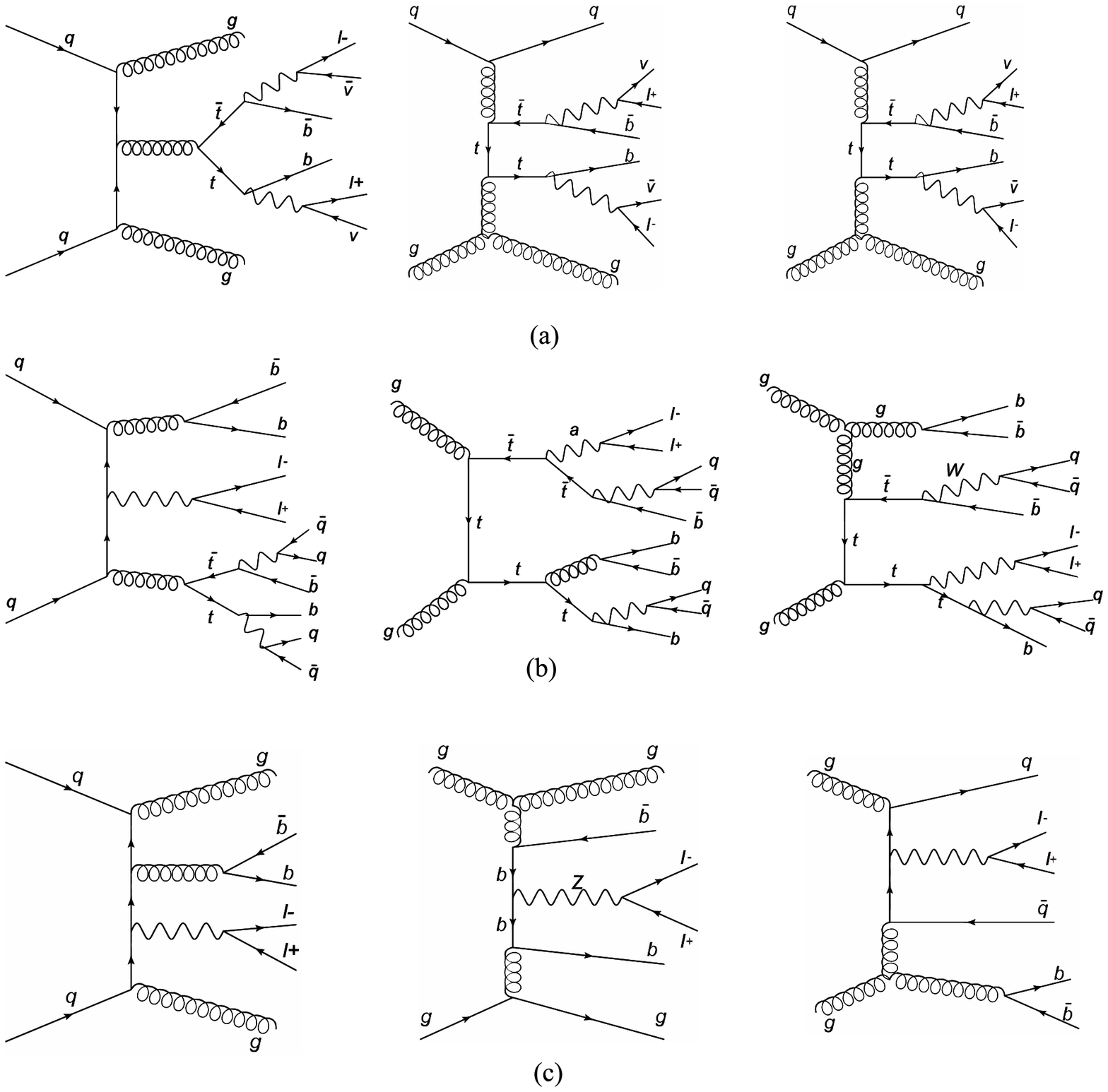}
\null\vspace{-0.2cm} \caption{Feynmann diagrams for typical RBs in
the case of 2-lepton mode.}
\label{double-leptons}
\end{figure}

\null\vspace{-0.5cm} To be able to show the $H$ resonance peak, we
consider the case of 2-lepton mode with the top quark pair from
$H$ decay decaying hadronically. When a top (anti-top) quark is
highly boosted, its decay products can be clustered into a single
fat jet by a clustering algorithm with a large $R$. And we can tag
such a top (anti-top) quark at a high efficiency using jet
substructure methods which are reviewed in paper
\cite{boosted-objects}\cite{jet-substructure}.

Some typical Feynman diagrams for S and IB before top quark decay
are shown in Fig.\,\ref{higgs-top-pair}. A signal event must
contains two opposite-sign leptons, two tagged b jets, and two
tagged t jets. The most important RBs which mimic the signal
are:\\
\null\noindent{\bf $\bm\bullet$ RB-1: $\bm t\bar {\bm
t}\bm{jj}$}\\
 Processes with a $t\bar{t}$ pair and two extra jets
in the final state. The $t\bar{t}$ pair decay semileptonically (cf. Fig.\,\ref{double-leptons}(a)).\\
\null\noindent{\bf$\bm\bullet$ RB-2: $\bm{\ell^+ \ell^- b}
\bar{\bm b}\bm{ jj}$}\\
 Processes with a $\ell^+ \ell^-$ pair, a
$b\bar{b}$ pair and two extra jets in the final state (cf. Fig\,
\ref{double-leptons}(c)). In {\bf RB-1} and {\bf RB-2}, we only
generate two large-$p^{}_T$ jets in the matrix element simulation
in order to simplify the calculation. A fully merged background
may modify the results, but the modification
will not be significant. \\
\null\noindent{\bf$\bm\bullet$ RB-3:
$\bm{\ell^+ \ell^- b}\bar{\bm
bt}\bar{\bm t}$}\\
 Processes with a $\ell^+\ell^-$ pair, a
$b\bar{b}$ pair and a $t\bar{t}$ pair in the final state. The
$t\bar{t}$ pair decay to jets (cf. Fig.\,\ref{double-leptons}(b)).

The simulation is similar to the case of the 3-lepton mode, while,
in addition, we should take the top quark tagging considering the
jet substructure. We make jet clustering using the
Cambridge-Aachen (CA) algorithm \cite{ca-jet} with radius $R=1.2$.
The jet pruning algorithm \cite{prune} with parameters $Z_{\rm{
cut}}$=0.1 and $RFactor_{\rm cut}$=0.5 is applied to the CA jets
for further suppressing RBs. If the transverse momentum of a
pruned CA jet is larger than 350 GeV, we retain it as a boosted
jet. The components of the CA jets with $p^{}_T<350$ GeV are then
clustered again by the anti-$k_T$ algorithm with $R=0.5$. A
b-tagging scheme with the same parameters as in
Ref.\,\cite{snowmass} is also performed. We then impose the
following cuts:

 \null\noindent{\bf Cut 1}: Requiring the event to contain two isolated
opposite-sign leptons and two tagged b jets, i.e.
\null\vspace{-0.2cm}
\begin{eqnarray}                              
N(\ell^+)=1,~N(\ell^-)=1,~~ N(b)=2.
\end{eqnarray}

\null\vspace{-1cm} \null\noindent{\bf Cut 2}: Let $j^{}_1$ be the
jet with largest $p^{}_T$ and $j^{}_2$ be the jet with the second
largest $p^{}_T$. We require
\begin{eqnarray}                            
p^{}_T(j^{}_i)>350 \,{\rm GeV},~~|\eta(j^{}_i)|<2
\end{eqnarray}
This can pick up the events containing two boosted jets.

\null\noindent{\bf Cut 3}: Requiring the mass of the two boosted
jets $j^{}_1$ and $j^{}_2$ to be in the neighborhood of $m^{}_t$,
i.e.
\begin{eqnarray}                          
150 \, {\rm GeV}<M(j^{}_i)<220 \, {\rm GeV}
\end{eqnarray}
This makes $j^{}_1$ and $j^{}_2$ two tagged boosted top jets.

\null\noindent{\bf Cut 4}: Since $H$ is produced by top-quark
fusion, its momentum should not be so large. When the two boosted
jets are the decay products of an s-channel $H$, the absolute
value of the vector sum of their 3-momenta should be relatively
small. So we require
\begin{eqnarray}                             
\left|\bm{p}(j^{}_1)+\bm{p}(j^{}_2)\right|<1 \,{\rm TeV}.
\end{eqnarray}
This makes the two boosted top jets coming from $H$ decay.

The cross sections (in fb) of S+IB, IB, and three RBs at the 50
TeV SppC after the cuts are listed in Table\,\ref{2Lefficiency}.
We see that after the four cuts, the backgrounds are reduced to
the same order of magnitude as S+IB. \null\vspace{-1.4cm}
\begin{widetext}

\begin{table}[h]                          
\caption{\label{2Lefficiency} Cross sections (in fb) of S+IB, IB
and RBs for various values of $M^{}_H$ at the 50 TeV SppC after
each cut in the case of 2-lepton mode.}
 \tabcolsep 12pt
\begin{tabular}{cccccccccc}
\hline \hline &&S+IB&&&IB&RB-1&RB-2&RB-3\\
              & 400 GeV  & 600 GeV & 800 GeV & 1 TeV &     &  &  &   \\
\hline
Cut 1 \& 2       &1.05     &0.70    &0.65    &0.62      &0.40   &34.6               &  1.79      & 0.06    \\
Cut 3            &0.086    &0.061   &0.063   &0.07      &0.036  &0.30               &  0.0096    & 0.00049 \\
Cut 4            &0.032    &0.022   &0.022   &0.026     &0.013  &0.067              &  0.0012    & 0.00011 \\
\hline \hline
\end{tabular}
\end{table}
\end{widetext}

\begin{figure}[h]
\includegraphics[width=0.5\textwidth]{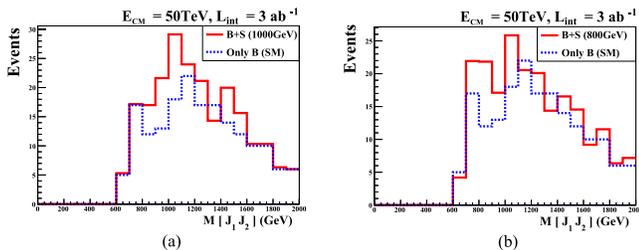}
\null\vspace{-0.8cm}
 \caption{\label{plot-MH}$M(j^{}_1,j^{}_2)$
distributions (red-solid) for an integrated luminosity of 3
ab$^{-1}$ at the 50 TeV SppC: (a) $M^{}_H=$ 1 TeV, (b)
$M^{}_H=800$ GeV. The SM background (blue-dotted) is also plotted
for comparison.}
\end{figure}

Since the two top jets from $H$ decay for $M^{}_H<800$ GeV can
hardly satisfy cut 2, we take the cases of $M^{}_H=1$ TeV and
$M^{}_H=800$ GeV to plot the invariant mass distributions for an
integrated luminosity of 3 ab$^{-1}$ at the 50 TeV SppC in
Fig.\,\ref{plot-MH}. the resonance peaks can be observed.
Note that the resonance peaks in Fig\,\ref{plot-MH} are not clear
enough for precision measurement of $M^{}_H$. A more
sophisticated, while complicated, top tagging algorithm
\cite{top-tagging-2008} \cite{top-tagging-2011} may improve the
results.

\null\noindent{\bf III. Summary}

 In this paper, we have studied the probe of
the gauge-phobic heavy Higgs bosons $H$, usually appear in new
physics models, at the 14 TeV LHC and the 50 TeV SppC via the
process $pp\to t\bar t t\bar t$ mainly contributed from gluon
fusion. We take the general (model-independent) $Ht\bar t$ Yukawa
interaction form in Eq.\,(\ref{Yukawa}).

Our calculation is at the hadron level including simulating the
jet formation and top quark tagging (with jet substructure) taking
account of the requirement of the detector acceptance. For
suppressing the SM background, we take two of the decay modes of
the top quarks including leptons, namely the 3-lepton mode and the
2-lepton mode. the former can have negligible SM RBs, and the
latter can make the resonance peak of $H$ visible.

In the 3-lepton mode, the needed integrated luminosities for
$1\sigma,3\sigma$ and $5\sigma$ deviations from the SM background
are shown in Table\,\ref{tripel-leptons-luminosity}. The case of
$5\sigma$ for $M^{}_H=600$ GeV, and the cases of $3\sigma$ and
$5\sigma$ for $M^{}_H\ge 800$ GeV at the LHC require very high
luminosities which need the upgraded high luminosity LHC. All
other cases can be reached at the present 14 TeV LHC and the SppC.
We have further shown that the mass of $H$ can be experimentally
determined by measuring the $p^{}_T(b_1)$-distribution at the 50
TeV SppC for an integrated luminosity of 3 ab$^{-1}$ (cf. Figs.\,
\ref{LHC-p_T} and \ref{SppC-p_T}).

In the 2-lepton mode, we imposed a series of cuts to suppress the
background, and extracted the process that the $t$ and $\bar t$
with hadronic decays are from the $H$ decay. The plotted invariant
mass distributions $M(j^{}_1,j^{}_2)$ at the 50 TeV SppC for an
integrated luminosity of 3 ab$^{-1}$ are shown in
Fig.\,\ref{plot-MH} which shows that the resonance peak of $H$ can
be observed. But the peak is not clear enough to give a very
accurate measurement for $M^{}_H$. More sophisticated top tagging
algorithm \cite{top-tagging-2008} \cite{top-tagging-2011} compared
with the simple one used here may help to highlight the resonance
peaks.

\null\noindent{\bf Acknowledgement} We would like to thank
Tsinghua National Laboratory for Information Science and
Technology for providing their computing facility. This work is
supported by the National Natural Science Foundation of China
under the grant numbers 11135003 and 11275102.

\bibliography{000}

\begin{thebibliography}{0}

\bibitem{ATLAS-CMS12}
G. Aad {\it et al.}, (ATLAS Collaboration),
\href{http://www.sciencedirect.com/science/article/pii/S037026931200857X}
{Phys. Lett. {\bf B 716} (2012) 1;}
W. Adam {\it et al.}, (CMS Collaboration),
\href{http://www.sciencedirect.com/science/article/pii/S0370269312008581}
{Phys. Lett. {\bf B 716} (2012) 30.}

\bibitem{CMS-JHEP13&ATLAS1305}
S. Chatrchyan {\it et al.} (CMS Collaboration),
\href{http://link.springer.com/article/10.1007/JHEP06\%282013\%29081}
{JHEP {\bf 1306} (2013) 81;}
S.M. Consonni {\it et al.} (ATLAS Collaboration),
\href{http://arxiv.org/abs/1305.3315}
{arXiv: 1305.3315.}

\bibitem{ATLAS-higgs-measurement}
ATLAS Collaboration,
\href{http://cds.cern.ch/record/1670012}
{ATLAS-CONF-2014-09 (2014).}

\bibitem{CMS-higgs-measurement}
CMS collaboration,
\href{http://cds.cern.ch/record/1728249}
{CMS-PAS-HIG-14-009 (2014).}

\bibitem{Flechl}
M. Flechl, on behalf of CMS and ATLAS Collaboration,
\href{http://arxiv.org/abs/1503.00632}
{arXiv:1503.00632.}

\bibitem{triviality}
R. Dashen and H. Neuberger,
\href{http://link.aps.org/doi/10.1103/PhysRevLett.50.1897}
{Phys. Rev. Lett. {\bf 50} (1983) 1897.}

\bibitem{unnaturalness}
L. Susskind,
\href{http://link.aps.org/doi/10.1103/PhysRevD.20.2619}
{Phys. Rev. D {\bf 20} (1979) 2619.}

\bibitem{vacuum-instability}
G. Degrassi, Giuseppe, et al.,
\href{http://link.springer.com/article/10.1007\%2FJHEP08\%282012\%29098}
{JHEP {\bf 1208} (2012) 098.}



\bibitem{ATLAS-PRD89}
G. Aad {\it et al.} (ATLAS Collaboration),
\href{http://journals.aps.org/prd/abstract/10.1103/PhysRevD.89.032002}
{Phys. Rev. D {\bf 89} (2014) 032002.}

\bibitem{KRX-PRD14}
Yu-Ping Kuang, Hong-Yu Ren, and Ling-Hao Xia,
\href{http://journals.aps.org/prd/abstract/10.1103/PhysRevD.90.115002}
{Phys. Rev, D {\bf 90} (2014) 115002.}

\bibitem{BhupalDev14}
P.S. Bhupal Dev and A. Pilaftsis,
\href{http://link.springer.com/article/10.1007\%2FJHEP12\%282014\%29024}
{JHEP {\bf 1412} (2014) 024.}

\bibitem{four-top-NLO}
G. Bevilacqua and M.Worek,
\href{http://link.springer.com/article/10.1007\%2FJHEP07\%282012\%29111}
{JHEP {\bf 1207} (2012) 111.}

\bibitem{madgraph5}
J. Alwall, M. Herquet, F. Maltoni, O.Mattelaer, and T. Stelzer,
\href{http://link.springer.com/article/10.1007\%2FJHEP06\%282011\%29128}
{JHEP {\bf 1106} (2011) 128.}

\bibitem{feynrules}
A.~Alloul, N.~D.~Christensen, C.~Degrande, C.~Duhr and B.~Fuks,
\href{http://www.sciencedirect.com/science/article/pii/S0010465514001350}
{Comput.Phys.Commun. {\bf 185} (2014).}

\bibitem{cteq6.1}
D. Stump, J. Huston, J. Pumplin, W.-K. Tung, H.-L. Lai, S.
Kuhlmann, and J. F. Owens,
\href{http://iopscience.iop.org/1126-6708/2003/10/046/}
{JHEP {\bf 0310} (2003) 046.}

\bibitem{pythia-6.4}
T. Sjostrand, S. Mrenna, and P. Skands,
\href{http://iopscience.iop.org/1126-6708/2006/05/026/}
{JHEP {\bf 0605} (2006) 026.}

\bibitem{delphes3}
J. Favereau, C. Delaere, P. Demin, A. Giammanco, V. Lemaitre, A. Mertens, and M. Selvaggi,
\href{http://link.springer.com/article/10.1007\%2FJHEP02\%282014\%29057}
{JHEP {\bf 1402} (2014) 057.}

\bibitem{fastjet}
M. Cacciari, G. P. Salam and G. Soyez,
\href{http://link.springer.com/article/10.1140\%2Fepjc\%2Fs10052-012-1896-2}
{Euro Phys. J. C {\bf 72} (2012) 1896.}

\bibitem{CMS-detector}
S. Chatrchyan, et al. (CMS collaboration),
\href{http://iopscience.iop.org/1748-0221/3/08/S08004}
{J. Instrumen. {\bf 3} (2008) S08004.}

\bibitem{CMS-TOP-13-012}
CMS Collaboration,
\href{https://cds.cern.ch/record/1644574}
{\newblock CERN-PH-EP-2014-222 (2014).}

\bibitem{ATLAS-4top}
Daniela. Paredes,
\href{http://cds.cern.ch/record/1627812}
{CERN-THESIS-2013-202.}

\bibitem{snowmass}
A. Avetisyan {\it et al.}, 
\href{http://arxiv.org/abs/1308.1636}
{arXiv:1308.1636 (2013).}

\bibitem{footnote-1}
Note that the width of the GPHB is not large, e.g., even for an
$M^{}_H=1$ TeV GPHB, its width is just 40 GeV. Thus in the
vicinity of $M^{}_H$, the contribution from the interference term
is much smaller than the total resonace contribution. So this
definition of $\sigma^{}_{\rm{ S}}$ essentially reflects the
property of the signal.

\bibitem{boosted-objects}
A. Abdesselam {\it et al.},
\href{http://link.springer.com/article/10.1140\%2Fepjc\%2Fs10052-011-1661-y}
{Eur. Phys. J. C {\bf 71} (2011) 1661.}

\bibitem{jet-substructure}
A. Altheimer {\it et al.},
\href{http://iopscience.iop.org/0954-3899/39/6/063001/}
{J. Phys. G {\bf 39} (2012) 063001.}

\bibitem{ca-jet}
Y. L. Dokshitzer, G. D. Leder, S. Moretti, and B. R. Webber.
\href{http://iopscience.iop.org/1126-6708/1997/08/001/}
{JHEP {\bf 9708} (1007) 001.}

\bibitem{prune}
S. D. Ellis, C. K. Vermilion and J. R. Walsh.
\href{http://journals.aps.org/prd/abstract/10.1103/PhysRevD.80.051501}
{Phys. Rev. D {\bf 80} (2009) 051501;}
\href{http://journals.aps.org/prd/abstract/10.1103/PhysRevD.81.094023}
{Phys. Rev. D {\bf 81} (2010) 094023.}

\bibitem{top-tagging-2008}
D. E. Kaplan, K. Rehermann, M. D. Schwartz, and B. Tweedie,
\href{http://journals.aps.org/prl/abstract/10.1103/PhysRevLett.101.142001}
{Phys.Rev. Lett. {\bf 101} (2008) 142001.}

\bibitem{top-tagging-2011}
T. Plehn and M. Spannowsky,
\href{http://iopscience.iop.org/0954-3899/39/8/083001}
{J. Phys. G {\bf 39} (2012) 083001.}

\end{thebibliography}

\end{document}